\documentclass[12pt]{article}

\renewcommand{\theequation}
{\arabic{section}.\arabic{equation}}

\makeatletter
\def\eqnarray{ \stepcounter{equation} \let\@currentlabel=\theequation
 \global\@eqnswtrue
 \global\@eqcnt\z@
 \tabskip\@centering
 \let\\=\@eqncr
 $$\halign to \displaywidth\bgroup\@eqnsel\hskip\@centering
 $\displaystyle\tabskip\z@{##}$&\global\@eqcnt\@ne
 \hfil$\displaystyle{{}##{}}$\hfil
 &\global\@eqcnt\tw@$\displaystyle\tabskip\z@{##}$\hfil
 \tabskip\@centering&\llap{##}\tabskip\z@\cr}
\makeatother

\makeatletter
\def\@arrayacol{\edef\@preamble{\@preamble \hskip .5\arraycolsep}}
\def\array{\let\@acol\@arrayacol \let\@classz\@arrayclassz
\let\@classiv\@arrayclassiv \let\\\@arraycr\def\@halignto{}\@tabarray}
\makeatother

\makeatletter
\newcounter{subeqncnt}
\def\thesubeqncnt{\alph{subeqncnt}}
\def\subequations{\begingroup%
   \stepcounter{equation}\edef\@tempa{\theequation}%
   \let\c@equation\c@subeqncnt\c@subeqncnt\z@
   \edef\theequation{\@tempa\noexpand\thesubeqncnt}}

\makeatother

\newcommand{\be}{\begin{equation}}
\newcommand{\ee}{\end{equation}}

\newcommand{\beqa}{\begin{eqnarray}}
\newcommand{\eeqa}{\end{eqnarray}}
\newcommand{\nn}{\nonumber}

\newcommand{\eqref}[1]{(\ref{#1})}

\def\CL {{\cal L}}

\begin{document}

\setlength{\baselineskip}{7mm}
\begin{titlepage}
\begin{flushright}
{\tt NRCPS-HE-57-07} \\

June, 2007
\end{flushright}

\vspace{1cm}

\begin{center}
{\it \Large Propagating Modes \\ of  \\
Non-Abelian Tensor Gauge Field of Second Rank
}

\vspace{1cm}

{ \it{Spyros Konitopoulos  } }
and
{ \it{George  Savvidy  } }

\vspace{0.5cm}

 {\it Institute of Nuclear Physics,} \\
{\it Demokritos National Research Center }\\
{\it Agia Paraskevi, GR-15310 Athens, Greece}

\end{center}

\vspace{1cm}

\begin{abstract}
In the recently proposed extension of the YM theory,
non-Abelian tensor gauge field of
the second rank is represented by a general tensor whose symmetric
part describes the propagation of charged gauge boson of helicity
two and its antisymmetric part - the helicity zero charged gauge
boson. On the non-interacting  level these polarizations are
similar to the polarizations of the graviton and of the Abelian
antisymmetric B field, but the interaction of these gauge bosons
carrying non-commutative internal charges cannot be
directly identified with the interaction of gravitons or B field.
Our intention here is
to illustrate this result from different perspectives which would
include Bianchi identity for the corresponding field strength
tensor  and the analysis of the second-order partial
differential equation which describes in this theory the
propagation of  non-Abelian tensor gauge field of the second rank.
Analyzing the interaction between two tensor currents caused  by
the exchange of these tensor gauge bosons we shall demonstrate
that the residue at the pole is the sum of three terms each of
which describes positive norm polarizations of helicities two and
zero bosons.

\end{abstract}

\end{titlepage}

\newpage

\pagestyle{plain}

\section{{\it Introduction}}

An infinite tower of massive particles of high spin naturally
appears in the spectrum of different string field theories
\cite{Witten:1985cc,Thorn:1985fa,Siegel:1985tw,Siegel:1988yz,Arefeva:1989cp}.
From the point of view of quantum field theory, string field
theories seem to contain an infinite number of nonrenormalizable
interactions between these fields, which are represented in the
string action by nonlocal cubic interaction terms
containing an exponential of a quadratic form in the momenta
\cite{Taylor:2003gn,Taylor:2006ye}.

It is generally expected that in the tensionless limit
or, what is equivalent, at high energy scattering
\cite{Gross:1987kz,Gross:1987ar,Gross} the string spectrum becomes
effectively massless and it is of great importance
to find out the corresponding action and its genuine symmetries
\cite{Witten:1988zd,Mende:1992pm,Mende:1989wt, Mende:1994wf, Moore:1993ns,
Giddings:2007bw,Savvidy:2003fx,Savvidy:2005fe}. On the quantum field theory
language this should be a field theory with infinite many
massless fields.

In quantum field theory the Lagrangian formulation of {\it free }
massless Abelian tensor gauge fields
has been  constructed in
 \cite{fierz,fierzpauli,yukawa1,wigner,schwinger,
Weinberg:1964cn,chang,singh,fronsdal,deWit:1979pe,Curtright:1980yk}.
The problem of introducing {\it interactions} appears
to be much more complex and there has been important progress
in defining self-interaction of higher-spin fields
in the light-cone formalism and
in the covariant formulation of the theories
\cite{Bengtsson:1983pd,Bengtsson:1983pg,Bengtsson:1986kh,berends,
Berends:1984wp,fronsdal2,Metsaev:2005ar,Boulanger:2006gr,Francia:2007qt,
Bekaert:2005vh,Engquist:2002vr,Sezgin:2001zs}.

In this respect it is appealing to extend the gauge principle
so that it will define the interaction of gauge fields which carry not only
non-commutative internal charges, but also arbitrary spins.
For that purpose it seems reasonable to define extended
non-Abelian gauge transformations acting on {\it charged tensor gauge fields}
and the corresponding field strength tensors,
which will enable the construction of a gauge invariant Lagrangian quadratic in field
strength tensors \cite{Savvidy:2005fi},
as it is the case in the Yang-Mills theory \cite{yang}.
The resulting gauge invariant Lagrangian will contain
kinetic terms for higher-spin fields and will uniquely define
their   cubic and quartic
interaction terms  \cite{Savvidy:2005fi,Savvidy:2005zm,Savvidy:2005ki}.

Here we shall follow the construction which is based on this
direct extension of non-Abelian gauge transformations
\cite{Savvidy:2005fi,Savvidy:2005zm,Savvidy:2005ki,Savvidy:2005at,Barrett:2007nn}.
Recall that non-Abelian gauge fields are defined as
rank-(s+1) tensor gauge fields
$
A^{a}_{\mu\lambda_1 ... \lambda_{s}}
$
\footnote{Tensor gauge fields
$A^{a}_{\mu\lambda_1 ... \lambda_{s}}(x),~~s=0,1,2,...$
are totally symmetric with respect to the
indices $  \lambda_1 ... \lambda_{s}  $. {\it A priori} the tensor fields
have no symmetries with respect to the first index  $\mu$.
In particular we have
$A^{a}_{\mu\lambda}\neq A^{a}_{\lambda\mu}$ and
$A^{a}_{\mu\lambda\rho}=A^{a}_{\mu\rho\lambda} \neq A^{a}_{\lambda\mu\rho}$.
The adjoint group index $a=1,...,N^2 -1$
for the $SU(N)$ gauge group. }
and that one can construct infinite series of forms
$ {{\cal L}}_{s}~( s=1,2,..)$ and ${{\cal L}}^{'}_{s}~( s=2,3,..)$ which are
invariant with respect to the gauge transformations.
These forms are quadratic in the field strength tensors
$G^{a}_{\mu\nu,\lambda_1 ... \lambda_s}$.  The construction of
invariant forms is based on the fact that
field strength tensors $G^{a}_{\mu\nu,\lambda_1 ... \lambda_s}$
transform   homogeneously
with respect to the gauge transformations.
Therefore the gauge invariant
Lagrangian describing dynamical tensor gauge bosons of all ranks
has the form \cite{Savvidy:2005fi,Savvidy:2005zm,Savvidy:2005ki}
\be\label{Lagrangian}
{\cal L} ~= ~
{{\cal L}}_1 +g_{2 }{{\cal L}}_2 +  g^{'}_{2 }{{\cal L}}^{'}_2 +....
\ee
where ${{\cal L}}_1$ is the Yang-Mills Lagrangian.
This Lagrangian contains {\it kinetic terms} of the tensor
gauge fields $A^{a}_{\mu},A^{a}_{\mu\lambda},.. $
and nonlinear terms which describe their interactions,
{\it cubic and quartic interactions} between lower- and higher-rank tensor gauge fields.
For the lower-rank tensor gauge fields the Lagrangian has the following form
\cite{Savvidy:2005fi,Savvidy:2005zm,Savvidy:2005ki}:
\beqa\label{totalactiontwo}
{{\cal L}}_1 &=& -{1\over 4}G^{a}_{\mu\nu}
G^{a}_{\mu\nu},\nn\\
{{\cal L}}_2  &=&-
{1\over 4}G^{a}_{\mu\nu,\lambda}G^{a}_{\mu\nu,\lambda}
-{1\over 4}G^{a}_{\mu\nu}G^{a}_{\mu\nu,\lambda\lambda}, \\
{{\cal L}}^{'}_2 &=
&+{1\over 4}G^{a}_{\mu\nu,\lambda}G^{a}_{\mu\lambda,\nu}
+{1\over 4}G^{a}_{\mu\nu,\nu}G^{a}_{\mu\lambda,\lambda}
+{1\over 2}G^{a}_{\mu\nu}G^{a}_{\mu\lambda,\nu\lambda},\nn
\eeqa
where generalized field strength tensors are:
\beqa
G^{a}_{\mu\nu} &=&
\partial_{\mu} A^{a}_{\nu} - \partial_{\nu} A^{a}_{\mu} +
g f^{abc}~A^{b}_{\mu}~A^{c}_{\nu}, \nn\\
G^{a}_{\mu\nu,\lambda} &=&
\partial_{\mu} A^{a}_{\nu\lambda} - \partial_{\nu} A^{a}_{\mu\lambda} +
g f^{abc}(~A^{b}_{\mu}~A^{c}_{\nu\lambda} + A^{b}_{\mu\lambda}~A^{c}_{\nu} ~),\nn\\
G^{a}_{\mu\nu,\lambda\rho} &=&
\partial_{\mu} A^{a}_{\nu\lambda\rho} - \partial_{\nu} A^{a}_{\mu\lambda\rho} +
g f^{abc}(~A^{b}_{\mu}~A^{c}_{\nu\lambda\rho} +
 A^{b}_{\mu\lambda}~A^{c}_{\nu\rho}+A^{b}_{\mu\rho}~A^{c}_{\nu\lambda}
 + A^{b}_{\mu\lambda\rho}~A^{c}_{\nu} ~).\nn
\eeqa
In the present paper we shall focus our attention on the lower-rank
tensor gauge field $A^{a}_{\mu\lambda}$, which in this theory is
a general nonsymmetric tensor with 4x4=16 space-time components
($A^{a}_{\mu\lambda}\neq A^{a}_{\lambda\mu}$ and it would have
$d \times d=d^2$  components in the d-dimensional
space-time)\footnote{One should multiply these numbers by the dimension
of the gauge group, $N^2 -1$ in the case of $SU(N)$.}.
It has been found that if
$g^{'}_{2} = g_{2}$, then the quadratic part of the Lagrangian
(\ref{totalactiontwo}), which defines the kinetic energy
of the field $A^{a}_{\mu\lambda}$,
has the following form \cite{Savvidy:2005fi,Savvidy:2005zm,Savvidy:2005ki}:
\be\label{thelagran}
 {\cal K}   =-{1\over
4}F^{a}_{\mu\nu,\lambda}F^{a}_{\mu\nu,\lambda} +
{1\over 4}F^{a}_{\mu\nu,\lambda}F^{a}_{\mu\lambda,\nu}+{1\over
4}F^{a}_{\mu\nu,\nu}F^{a}_{\mu\lambda,\lambda},\nn
\ee
where
\be\label{abelianpart}
F^{a}_{\mu\nu,\lambda}=
\partial_{\mu} A^{a}_{\nu\lambda} - \partial_{\nu} A^{a}_{\mu\lambda},
\ee
and describes the propagation of {\it helicity-two and helicity-zero
$\lambda = \pm 2, 0$ charged gauge bosons}.
This means  that within the 16 fields of nonsymmetric tensor gauge
field $A^{a}_{\mu\lambda}$ only {\it three positive norm polarizations}
are propagating and that the rest of them are pure gauge fields.
On the non-interacting  level, when we consider only the
kinetic term (\ref{thelagran}) of the full Lagrangian (\ref{Lagrangian}),
these polarizations are similar to the polarizations of the
graviton and of the Abelian antisymmetric B field
\cite{Kalb:1974yc,Cremmer:1973mg,Townsend:1976vc,Nambu:1975ba,Einstein1950},
but the interaction of these gauge bosons
carrying non-commutative internal charges is
uniquely defined by the full Lagrangian (\ref{Lagrangian}) and cannot be
directly identified with the interactions of gravitons or B field
\cite{Savvidy:2005fi,Savvidy:2005zm,Savvidy:2005ki}.

Our intention here is to illustrate this result from different
perspectives which would include Bianchi identity for the field
strength tensor $F^{a}_{\mu\nu,\lambda}$ and the direct analysis
of the second-order partial differential equation which describes
in this theory the propagation of the free tensor gauge field
$A^{a}_{\mu\lambda}$. This equation follows from the variation of
the Lagrangian (\ref{thelagran}) with respect to $A^{a}_{\nu\lambda}$
\cite{Savvidy:2005fi,Savvidy:2005zm,Savvidy:2005ki}.
With this in mind  we shall present a general
method for counting the propagating modes in  gauge field theories
and apply it to the tensor gauge field theory in order to
illustrate the fact that the second-rank tensor
gauge field $A^{a}_{\mu\lambda}$ describes three polarizations
$\lambda =\pm 2,0$ of massless charged tensor gauge bosons.

We shall see that the gauge invariance, together with the geometrical
properties of the field strength tensor, poses strong
restrictions on the divergences of the currents.
 Analyzing the interaction between two tensor currents caused  by the
exchange of these tensor gauge bosons we shall demonstrate that
the residue at the pole is the sum of three terms each of which
describes positive norm polarizations of helicities $\lambda =\pm 2,0$.

\section{{\it Gauge Symmetries and Current Conservation}}

Pure kinetic term of the
Lagrangian (\ref{totalactiontwo}), (\ref{thelagran}), which
describes the propagation of the tensor gauge field
($A^{a}_{\mu\lambda}\neq A^{a}_{\lambda\mu}$),
has the following form \cite{Savvidy:2005fi,Savvidy:2005zm,Savvidy:2005ki}:
\be
\label{kineticterm}
{\cal K}  =-{1\over
4}F^{a}_{\mu\nu,\lambda}F^{a}_{\mu\nu,\lambda} +
{1\over 4}F^{a}_{\mu\nu,\lambda}F^{a}_{\mu\lambda,\nu}+{1\over
4}F^{a}_{\mu\nu,\nu}F^{a}_{\mu\lambda,\lambda}
\ee
and is invariant with respect to a pair of {\it complementary} gauge transformations
$\delta$ and $\tilde{\delta}$. When the coupling constant $g$ is taken to
vanish, these gauge transformations reduce to the following form
\cite{Savvidy:2005ki}:
\be\label{polygauge}
\delta  A^{a}_{\mu\lambda } =   \partial_{\mu} \xi^{a}_{\lambda }
\ee
and
\be\label{freedoublepolygaugesymmetric}
\tilde{\delta}  A^{a}_{\mu\lambda } =   \partial_{\lambda }
\eta^{a}_{\mu}.
\ee
The field strength tensor $F^{a}_{\mu\nu,\lambda}$ (\ref{abelianpart})
transforms with respect to these transformations as follows:
\beqa\label{fieldstrengthtrans}
\delta_{\xi} F^{a}_{\mu\nu,\lambda } &=& 0 , \nn\\
\tilde{\delta}_{\eta} F^{a}_{\mu\nu,\lambda } &=&   \partial_{\lambda}
(\partial_{\mu} \eta^{a}_{\nu} - \partial_{\nu} \eta^{a}_{\mu}).
\eeqa
Therefore the kinetic term ${\cal K} $ is obviously invariant with respect to the
first group of gauge transformations  $\delta$ and we have $\delta {\cal K} =0$,
but it is less trivial to see that it is
also invariant with respect to the complementary gauge transformations
$\tilde{\delta}$. The $\tilde{\delta}$ transformation of ${\cal K}$ is
\beqa
\tilde{\delta}{\cal K} &=&-{1\over 2}F^{a}_{\mu\nu,\lambda}
\partial_{\lambda}(\partial_{\mu}\eta^{a}_{\nu}
-\partial_{\nu}\eta^{a}_{\mu}  ) +
{1\over 2}F^{a}_{\mu\nu,\lambda}\partial_{\nu}(\partial_{\mu}\eta^{a}_{\lambda}
-\partial_{\lambda}\eta^{a}_{\mu}  ) + {1\over
2}F^{a}_{\mu\nu,\nu}\partial_{\lambda}(\partial_{\mu}\eta^{a}_{\lambda}
-\partial_{\lambda}\eta^{a}_{\mu}  )=\nn\\
&=&{1\over 2}F^{a}_{\mu\nu,\lambda}  \partial_{\lambda}
\partial_{\nu}\eta^{a}_{\mu}+ {1\over
2}F^{a}_{\mu\nu,\nu}\partial_{\lambda}(\partial_{\mu}\eta^{a}_{\lambda}
-\partial_{\lambda}\eta^{a}_{\mu}  ), \nn
\eeqa
where we combined the first, the second and the forth terms and used the fact that
the third term is identically equal to zero. Just from the symmetry properties
of the field strength tensor it is not obvious to see why the rest of the terms
are equal to zero. Therefore we shall use the explicit form of the field strength
tensor $F^{a}_{\mu\nu,\lambda}$, which gives
\beqa
\tilde{\delta}{\cal K}
&=&{1\over 2}(\partial_{\mu} A^{a}_{\nu\lambda} - \partial_{\nu} A^{a}_{\mu\lambda}) \partial_{\lambda}
\partial_{\nu}\eta^{a}_{\mu}+ {1\over
2}(\partial_{\mu} A^{a}_{\nu\nu} - \partial_{\nu} A^{a}_{\mu\nu})
\partial_{\lambda}(\partial_{\mu}\eta^{a}_{\lambda}
-\partial_{\lambda}\eta^{a}_{\mu}  ). \nn
\eeqa
From the corresponding action $S_0 =  \int  {\cal K}dx$ after partial differentiation
we shall get that the term $\partial_{\mu} A^{a}_{\nu\nu} ~\cdot~
\partial_{\lambda}(\partial_{\mu}\eta^{a}_{\lambda}
-\partial_{\lambda}\eta^{a}_{\mu}  )$ gives a zero contribution and the rest of
the terms cancel each other $\int ({1\over 2}(\partial_{\mu} A^{a}_{\nu\lambda}
- \partial_{\nu} A^{a}_{\mu\lambda}) \cdot \partial_{\lambda}
\partial_{\nu}\eta^{a}_{\mu} - {1\over
2} \partial_{\nu} A^{a}_{\mu\nu} \cdot
\partial_{\lambda}(\partial_{\mu}\eta^{a}_{\lambda}
-\partial_{\lambda}\eta^{a}_{\mu}  ))dx =0$.
This demonstrates the invariance  of the action with respect to the
$\delta$ and $\tilde{\delta}$ transformations defined by (\ref{polygauge}) and
(\ref{freedoublepolygaugesymmetric}) when $g_2 = g^{'}_{2}$ in
(\ref{Lagrangian}) \cite{Savvidy:2005fi}.

Let us now consider the interaction of the tensor gauge field $A^{a}_{\mu\lambda}$
with the current $J^{a}_{\mu\nu}$, as it is defined by the
Lagrangian (\ref{Lagrangian}) ($J^{a}_{\mu\nu} \neq J^{a}_{\nu\mu}$).
In order to see what type of restrictions are imposed on the current
by the gauge symmetries of the action we shall consider
the equation of motion which follows from the Lagrangian (\ref{Lagrangian})
( see \cite{Savvidy:2005fi,Savvidy:2005zm,Savvidy:2005ki} for details):
\beqa\label{freeequations}
\partial_{\mu} F^{a}_{\mu\nu,\lambda}
-{1\over 2} (\partial_{\mu} F^{a}_{\mu\lambda,\nu}
+\partial_{\mu} F^{a}_{\lambda\nu,\mu}
+\partial_{\lambda}F^{a}_{\mu\nu,\mu}
+\eta_{\nu\lambda} \partial_{\mu}F^{a}_{\mu\rho,\rho}) = J^{a}_{\nu\lambda}.
\eeqa
This equation contains two terms  $\partial_{\mu} F^{a}_{\mu\nu,\lambda}$
and
$-{1\over 2} (\partial_{\mu} F^{a}_{\mu\lambda,\nu}
+\partial_{\mu} F^{a}_{\lambda\nu,\mu}
+\partial_{\lambda}F^{a}_{\mu\nu,\mu}
+\eta_{\nu\lambda} \partial_{\mu}F^{a}_{\mu\rho,\rho})$,
which arise from ${{\cal L}}_2$ and ${{\cal L}}^{'}_2$ respectively.
The derivatives  over $\partial_{\nu}$ of both terms in the equation
are equal to zero separately.
Indeed, due to the antisymmetric properties of the field strength
tensor $F^{a}_{\mu\nu,\lambda}$ under exchange of $\mu$ and $\nu$
we have
$$
\partial_\nu \partial_{\mu} F^{a}_{\mu\nu,\lambda} =0,
$$
as well as
$$-{1\over 2} \partial_\nu \{\partial_{\mu} F^{a}_{\mu\lambda,\nu}
+\partial_{\mu} F^{a}_{\lambda\nu,\mu}
+\partial_{\lambda}F^{a}_{\mu\nu,\mu}
+\eta_{\nu\lambda} \partial_{\mu}F^{a}_{\mu\rho,\rho}\}=0.
$$
Thus it follows from (\ref{freeequations}) that
\be\label{divergentfirst}
\partial_{\nu} J^{a}_{\nu\lambda}=0.
\ee
Hence, the current $J^{a}_{\nu\lambda}$ must be divergenceless over
its first index. Now let us take derivative over
$\partial_\lambda$ of the left-hand side of the equation (\ref{freeequations}),
that is, the derivative over the second index of the
nonsymmetric current $J^{a}_{\nu\lambda}$.
We see that
$$
\partial_\lambda \partial_{\mu} F^{a}_{\mu\nu,\lambda} \neq 0,
$$
as well as
$$
-{1\over 2} \partial_\lambda \{\partial_{\mu} F^{a}_{\mu\lambda,\nu}
+\partial_{\mu} F^{a}_{\lambda\nu,\mu}
+\partial_{\lambda}F^{a}_{\mu\nu,\mu}
+\eta_{\nu\lambda} \partial_{\mu}F^{a}_{\mu\rho,\rho} \} =
-{1\over 2} \partial_\lambda \{
\partial_{\mu} F^{a}_{\lambda\nu,\mu}
+\partial_{\lambda}F^{a}_{\mu\nu,\mu}
+\eta_{\nu\lambda} \partial_{\mu}F^{a}_{\mu\rho,\rho} \}\neq0.
$$
Thus, is not obvious to see the conservation of the
nonsymmetric current $J^{a}_{\nu\lambda}$ with respect
to its second index $\lambda$ alone from the antisymmetric properties
of the field strength tensor. Therefore we have to use the explicit
form of the field strength tensor $F^{a}_{\mu\nu,\lambda} =
\partial_{\mu} A^{a}_{\nu \lambda} - \partial_{\nu} A^{a}_{\mu \lambda}$, this gives
\beqa
\partial_\lambda \partial_{\mu} F^{a}_{\mu\nu,\lambda}
-{1\over 2} \partial_\lambda \{ \partial_{\mu} F^{a}_{\mu\lambda,\nu}
+\partial_{\mu} F^{a}_{\lambda\nu,\mu}
+\partial_{\lambda}F^{a}_{\mu\nu,\mu}
+\eta_{\nu\lambda} \partial_{\mu}F^{a}_{\mu\rho,\rho} \}=\nn\\
=\partial_\lambda \partial_{\mu} F^{a}_{\mu\nu,\lambda}
-{1\over 2} \partial_\lambda \partial_{\mu} F^{a}_{\lambda\nu,\mu}
-{1\over 2} \partial^2 F^{a}_{\mu\nu,\mu}
-{1\over 2} \partial_\nu   \partial_{\mu}F^{a}_{\mu\rho,\rho} =\nn\\
= {1\over 2} \partial_\lambda \partial_{\mu}
(\partial_{\mu}  A^{a}_{\nu\lambda} -\partial_{\nu}  A^{a}_{\mu\lambda})
-{1\over 2} \partial^2
(\partial_{\mu}  A^{a}_{\nu\mu} -\partial_{\nu}  A^{a}_{\mu\mu})
-{1\over 2} \partial_\nu   \partial_{\mu}
(\partial_{\mu}  A^{a}_{\rho\rho} -\partial_{\rho}  A^{a}_{\mu\rho}) =0.\nn
\eeqa
Therefore the sum of the two nonzero
expressions presented above are equal to zero,
thus \cite{Savvidy:2005fi,Savvidy:2005zm,Savvidy:2005ki}
\be\label{divergentsecond}
\partial_{\lambda} J^{a}_{\nu\lambda}=0.
\ee
The natural question which arises here is connected with the fact that
in order to see these cancellations one should use the explicit form of the field
strength tensor $F^{a}_{\mu\nu,\lambda}$, and it remains a
mystery, why this takes place only when the relative
coefficient between the invariant forms  ${\cal L}_{2}$
and $ {\cal L}_{2}^{'}$ is equal to one ($g_2 = g^{'}_{2}$ in
(\ref{Lagrangian})) \cite{Savvidy:2005fi}. Our main concern
therefore is to understand  the general reason for these
cancellations without referring to the explicit
form of the field strength tensor.
As we shall see, the deep reason for this cancellations
lies is the Bianchi identity (\ref{newbianchi}), (\ref{Bianchiidentity1})
for the field strength tensor
\be
\label{introductionintoBianchiidentity}
\partial_{\mu}F_{\nu\lambda,\rho}+\partial_{\nu}F_{\lambda\mu,\rho}
+\partial_{\lambda}F_{\mu\nu,\rho}=0,
\ee
which we shall derive in the Appendix A.
Indeed, we can evaluate the derivative of the l.h.s of the equation
(\ref{freeequations}) to the following form:
\beqa\label{paradigma}
\partial_\lambda\{\partial_{\mu}F_{\mu\nu,\lambda}-{1\over 2} (\partial_{\mu}
F^{a}_{\mu\lambda,\nu}
+\partial_{\mu} F^{a}_{\lambda\nu,\mu}
+\partial_{\lambda}F^{a}_{\mu\nu,\mu}
&+&\eta_{\nu\lambda} \partial_{\mu}F^{a}_{\mu\rho,\rho})\}=\\
=&-&{1\over 2}\{\partial^{2}F_{\mu\nu,\mu}+\partial_{\mu}\partial_{\nu}
F_{\mu\rho,\rho}+\partial_{\mu}\partial_{\lambda}F_{\nu\lambda,\mu} \},\nn
\eeqa
where we have used only the antisymmetric property of
$F_{\mu\nu,\lambda}$ to cancel
the second term and to combine the first one with the third one of the l.h.s of the
above equation. Now, we shall take advantage of the Bianchi identity.
Taking the derivative of the Bianchi identity
(\ref{introductionintoBianchiidentity}) over $\partial_\mu$
and setting $\nu=\rho$ we  get
\be\label{Bianchideriv}
\partial^{2}F_{\mu\nu,\mu}+\partial_{\mu}\partial_{\nu}
F_{\mu\rho,\rho}+\partial_{\mu}\partial_{\lambda}F_{\nu\lambda,\mu} \equiv 0
\ee
and can clearly see that the last expression in  (\ref{paradigma}) coincides
with the left-hand side of this contracted Bianchi identity
and is therefore equal to zero.
Thus (\ref{divergentsecond}) holds -
$
\partial_{\lambda} J^{a}_{\nu\lambda}=0.
$

In other words, if one repeats these calculations for arbitrary
coefficients $g_2$   and  $g^{'}_{2}$  in the Lagrangian (\ref{Lagrangian})
$g_2{\CL}_2 + g^{'}_{2}{\CL}^{'}_2$,
then the last expression in parenthesis (\ref{paradigma}) will take the form
$$
\partial^{2}F_{\mu\nu,\mu} + \partial_{\mu}\partial_{\nu}
F_{\mu\rho,\rho} + (2~{g_2 \over g^{'}_{2}}-1)
\partial_{\mu}\partial_{\lambda}F_{\nu\lambda,\mu}.
$$
Comparing it with Bianchi identity (\ref{Bianchideriv}) one can see that
it is equal to zero only if $g_2=g^{'}_{2}$
and therefore only in that case (\ref{divergentsecond}) holds.

It seems that this situation
is similar to that in gravity, where
both tensors $R_{\mu\nu}$
and $g_{\mu\nu} R$  have correct transformation properties and therefore
can be present in the equation of motion \cite{Einstein1915}
\be
R_{\mu\nu} - c~ g_{\mu\nu} R = T_{\mu\nu}
\ee
with unknown coefficient $c$, but the Bianci  identity
$R^{\mu}_{\nu;\mu} - (1/2)R_{;\nu}=0 $ tells that the coefficient $c$
should be taken equal to $1/2$  \cite{Einstein1950}.

The total conservation of nonsymmetric ($J^{a}_{\mu\nu} \neq J^{a}_{\nu\mu}$)
current
(\ref{divergentfirst})
and (\ref{divergentsecond})
$$
\partial_{\nu} J^{a}_{\nu\lambda}=0,~~~~~\partial_{\lambda} J^{a}_{\nu\lambda}=0
$$
is adequate to cast the theory unitary at free level. In the next
section we shall present a general method for counting the
propagating modes in gauge field theories and shall see how the gauge
invariance guarantees the elimination of all negative-norm states.

\section{{\it Counting Propagating Modes}}

As we have seen above, the equation of motion (\ref{freeequations})
which describes the
propagation of the second-rank tensor gauge field
($A^{a}_{\mu\lambda}\neq A^{a}_{\lambda\mu}$) has the following
form\footnote{The Lorentz indexes of the tensor fields
are raised and lowered with flat metric
$\eta_{\mu\nu}=(-1,1,1,1) $.}
\cite{Savvidy:2005fi,Savvidy:2005zm,Savvidy:2005ki}:
\beqa\label{freeequations1}
\partial^{\mu} F^{a}_{\mu\nu,\lambda}
-{1\over 2} (\partial^{\mu} F^{a}_{\mu\lambda,\nu}
+\partial^{\mu} F^{a}_{\lambda\nu,\mu}
+\partial_{\lambda}F^{a~~\mu}_{\mu\nu,}
+\eta_{\nu\lambda} \partial^{\mu}F^{a~~\rho}_{\mu\rho,}) = J^{a}_{\nu\lambda},
\eeqa
where
$
F^{a}_{\mu\nu,\lambda}=
\partial_{\mu} A^{a}_{\nu\lambda} - \partial_{\nu} A^{a}_{\mu\lambda}
$
and is invariant with respect to a pair of complementary gauge
transformations $\delta$ (\ref{polygauge}) and $\tilde{\delta}$
(\ref{freedoublepolygaugesymmetric})
\be\label{dualgaugetransfromations}
 A^{a}_{\mu\lambda } ~~ \rightarrow~~ A^{a}_{\mu\lambda } ~+~
  \partial_{\mu} \xi^{a}_{\lambda }+ \partial_{\lambda }
\eta^{a}_{\mu}
\ee
with arbitrary gauge parameters $\xi^{a}_{\lambda }$ and $\eta^{a}_{\mu}$.
The equivalent form of the equations of motion (\ref{freeequations})
in terms of the gauge field is
\cite{Savvidy:2005fi,Savvidy:2005zm,Savvidy:2005ki}
\beqa\label{mainequation}
\partial^{2}(A^{a}_{\nu\lambda} -{1\over 2}A^{a}_{\lambda\nu})
-\partial_{\nu} \partial^{\mu}  (A^{a}_{\mu\lambda}-
{1\over 2}A^{a}_{\lambda\mu} )&-&
\partial_{\lambda} \partial^{\mu}  (A^{a}_{\nu\mu} - {1\over 2}A^{a}_{\mu\nu} )
+\partial_{\nu} \partial_{\lambda} ( A^{a~\mu}_{\mu}-{1\over 2}A^{a~\mu}_{\mu})+\nn\\
&+&{1\over 2}\eta_{\nu\lambda} ( \partial^{\mu}
\partial^{\rho}A^{a}_{\mu\rho} -  \partial^{2}A^{a~\mu}_{\mu})=J^{a}_{\nu\lambda}.
\eeqa
In momentum space this type of second-order partial differential
equations can always   be
represented as matrix equation of the following general form
\be\label{tensorfieldequation}
H_{\alpha\acute{\alpha}}^{~~~\gamma\acute{\gamma}}(k)~
A^{a}_{\gamma\acute{\gamma}}= J^{a}_{\alpha\acute{\alpha}},
\ee
where $H_{\alpha\acute{\alpha}}^{~~~\gamma\acute{\gamma}}(k)$ is a matrix operator
quadratic in momentum $k_{\mu}$. In our case it has the following form
\cite{Savvidy:2005fi,Savvidy:2005zm,Savvidy:2005ki}:
\beqa
\label{quadraticform}
H_{\alpha\acute{\alpha}\gamma\acute{\gamma}}(k)=
(-\eta_{\alpha\gamma}\eta_{\acute{\alpha}\acute{\gamma}}
+{1 \over 2}\eta_{\alpha\acute{\gamma}}\eta_{\acute{\alpha}\gamma}
+{1 \over 2}\eta_{\alpha\acute{\alpha}}\eta_{\gamma\acute{\gamma}})k^2
+\eta_{\alpha\gamma}k_{\acute\alpha}k_{\acute{\gamma}}
+\eta_{\acute\alpha \acute{\gamma}}k_{\alpha}k_{\gamma}\nn\\
-{1 \over 2}(\eta_{\alpha\acute{\gamma}}k_{\acute\alpha}k_{\gamma}
+\eta_{\acute\alpha\gamma}k_{\alpha}k_{\acute{\gamma}}
+\eta_{\alpha\acute\alpha}k_{\gamma}k_{\acute{\gamma}}
+\eta_{\gamma\acute{\gamma}}k_{\alpha}k_{\acute\alpha}),
\eeqa
with the property that $H_{\alpha\acute{\alpha}\gamma\acute{\gamma}}=
H_{\gamma\acute{\gamma}\alpha\acute{\alpha}}$.
First of all we shall solve the equation in the case when there are no
interactions,  $J^{a}_{\alpha\acute{\alpha}}=0$:
\be\label{basicequation}
H_{\alpha\acute{\alpha}}^{~~~\gamma\acute{\gamma}}(k)~
A^{a}_{\gamma\acute{\gamma}} =0.
\ee
The vector space of independent solutions
$A_{\gamma\acute{\gamma}}$   of this system of
equations crucially depends on the rank  of the matrix
$H_{\alpha\acute{\alpha}}^{~~~\gamma\acute{\gamma}}(k)$. If the matrix
operator $H$ has dimension $d \times d$ and its rank  is $rankH=r$,
then the vector space has the dimension
$$
{\cal N}= d-r.
$$
Because the matrix operator $H_{\alpha\acute{\alpha}}^{~~~\gamma\acute{\gamma}}(k)$
explicitly depends on the momentum $k_{\mu}$ its $rankH=r$ also depends on
momenta and therefore the number of independent solutions
${\cal N}$  depends on momenta
\be
{\cal N}(k) = d-r(k)~.
\ee
Analyzing the $rankH$ of the matrix operator $H$
one can observe that it depends on the value of momentum
square $k_{\mu}^2$. When $k_{\mu}^2 \neq 0$ - off mass-shell momenta - the
vector space consists of {\it pure gauge fields}. When $k_{\mu}^2 = 0$ - on mass-shell
momenta - the vector space consists of {\it pure gauge fields and propagating
modes}. Therefore the number of propagating modes can be calculated from
the following relation:
\be
\sharp ~ of~ propagating ~modes= {\cal N}(k)\vert_{k^2 = 0} -
{\cal N}(k)\vert_{k^2 \neq 0}=
 rankH \vert_{k^2\ne 0}-rankH \vert_{k^2=0}.
\ee
Before considering the equation of motion for the
tensor gauge field  (\ref{basicequation}),
let us consider for illustration some important examples.

\subsection{\it Vector Gauge Field}

The kinetic term of Lagrangian which describes the propagation
of free vector gauge field
is
\be
{{\cal K}}=-{1\over 4}F_{\mu\nu}F^{\mu\nu}
\ee
and the corresponding equation of
motion in momentum space is
\beqa\label{qedeqs}
H_{\alpha}^{~~\gamma}e_{\gamma}=(-k^{2}\delta_{\alpha}^{\gamma}+k_{\alpha}k^{\gamma})e_{\gamma}=
0,
\eeqa
where $A_{\mu} = e_{\mu} \exp {(ikx)}$. We can always choose the momentum vector
in the third direction  $k^{\mu} = (\omega, 0,0,k)$ and the matrix operator
$H$ takes the form
$$
H_{\alpha}^{~~\gamma}= \left(\begin{array}{cccc}
-k^{2}&0&0& -k \omega  \\
0&\omega^2 -k^2  &0&0 \\
0&0&\omega^2 -k^2 &0 \\
k \omega &0&0&\omega^{2} \\
\end{array} \right).
$$
If $\omega^2 - k^2 \neq 0$, the rank of the 4-dimensional matrix
$H_{\alpha}^{~\gamma}$ is $rankH\vert_{\omega^2 - k^2 \ne 0}=3$
and the number of independent solutions is 4-3=1. As one can see
from the relation  $H_{\alpha}^{~~\gamma}(k)k_{\gamma}=0$ this
solution is proportional to the momentum $e_{\mu}=
k_{\mu}=(-\omega, 0,0,k)$ and is a pure-gauge field. This is a
consequence of the gauge invariance of the theory
$e_{\mu}\rightarrow e_{\mu} + a k_{\mu}$. If $\omega^2 - k^2 = 0$,
then the rank of the matrix drops, $rankH\vert_{\omega^2 - k^2 =
0}=1$, and the number of independent solutions increases: 4-1=3.
These three solutions of equations (\ref{qedeqs}) are
$$
e_{\gamma}^{(gauge)}={1\over\sqrt{2}}\left(\begin{array}{cccc}
-1 \\
0 \\
0 \\
1 \\
\end{array} \right),~~ e_{\gamma}^{(1)}=\left(\begin{array}{cccc}
0 \\
1 \\
0 \\
0 \\
\end{array} \right),~~e_{\gamma}^{(2)}= \left(\begin{array}{cccc}
0 \\
0 \\
1 \\
0 \\
\end{array} \right),
$$
from which the first one is a pure-gauge field ($\sim k_{\gamma}$),
while the remaining two are the physical modes, perpendicular to the direction
of the wave propagation.
The general solution at
$\omega^2 - k^2 = 0$
will be a linear combination of these three eigenvectors:
$$e_{\gamma}=ak_{\gamma}+c_{1}e_{\gamma}^{(1)}+c_{2}e_{\gamma}^{(2)},$$
where $a,c_{1},c_{2}$ are arbitrary constants.
We see that the number of propagating modes  is
$$
rankH\vert_{\omega^2 - k^2 \ne 0}-
rankH\vert_{\omega^2 - k^2 = 0}=3-1=2,
$$
as it should be.

\subsection{\it Symmetric Tensor Gauge Field }

The free gravitational field is described in terms of a symmetric
second-rank tensor field $h_{\mu\nu}$ and is governed by the
Einstein and Pauli-Fierz equation:
\be\label{einstein}
\partial^{2} h_{\nu\lambda}
-\partial_{\nu} \partial^{\mu}  h_{\mu\lambda} -
\partial_{\lambda} \partial^{\mu}  h_{\mu\nu}
+ \partial_{\nu} \partial_{\lambda}  h^{~\mu}_{\mu} +\eta_{\nu\lambda}
(\partial^{\mu} \partial^{\rho}h_{\mu\rho} - \partial^{2}
 h^{~\mu}_{\mu}) =0,
\ee
which is invariant with respect to  the gauge transformations
\beqa
\label{quagetransfgravity}
\delta
h_{\mu\lambda}=\partial_{\mu}\xi_{\lambda}+\partial_{\lambda}\xi_{\mu},
\eeqa
which preserve the symmetry properties of $A_{\mu\nu}$. The corresponding
matrix operator is:
\beqa\label{eqsmatrixgravity}
H_{\alpha\acute{\alpha}\gamma\acute{\gamma}}(k)&=&\{\eta_{\alpha\acute{\alpha}}
\eta_{\gamma\acute{\gamma}}-{1\over 2}
(\eta_{\alpha\gamma}\eta_{\acute{\alpha}\acute{\gamma}}+
\eta_{\alpha\acute{\gamma}}\eta_{\acute{\alpha} \gamma})\}
k^{2}
-\eta_{\alpha\acute{\alpha}}k_{\gamma}k_{\acute{\gamma}}-
\eta_{\gamma\acute{\gamma}}k_{\alpha}k_{\acute{\alpha}}+\nn\\
&+&{1\over 2}(\eta_{\alpha\acute{\gamma}}k_{\gamma}k_{\acute{\alpha}}+
\eta_{\acute{\alpha}\acute{\gamma}}k_{\alpha}k_{\gamma}
+\eta_{\alpha\gamma}k_{\acute{\alpha}}k_{\acute{\gamma}}+
\eta_{\acute{\alpha}\gamma}k_{\alpha}k_{\acute{\gamma}})
\eeqa
and is a $10 \times 10$ matrix in four-dimensional space-time with the
property $H_{\alpha\acute{\alpha}\gamma\acute{\gamma}}=
H_{\acute{\alpha}\alpha\gamma\acute{\gamma}} =
H_{\alpha\acute{\alpha}\acute{\gamma}\gamma}=
H_{\gamma\acute{\gamma}\alpha\acute{\alpha}}$ and is presented in
Appendix B.

If $\omega^2 - k^2 \neq 0$, the rank of the 10-dimensional
matrix
$
H_{\alpha\acute{\alpha}}^{~~~\gamma\acute{\gamma}}(k)
$
is equal to $rankH\vert_{\omega^2 - k^2 \ne 0}=6$
and the number of independent solutions is $10-6=4$.
These four symmetric solutions are pure-gauge tensor fields. Indeed,
if again we choose the coordinate system so that
$k^{\gamma}=(\omega,0,0,k)$, then
one can find the following four linearly independent solutions:
\beqa
e_{\gamma\acute{\gamma}}=
\left(\begin{array}{cccc}
-\omega^{2}&0&0&0 \\
0&0&0&0 \\
0&0&0&0 \\
0&0&0&k^{2} \\
\end{array} \right),
\left(\begin{array}{cccc}
0&-\omega&0&0 \\
-\omega&0&0&k \\
0&0&0&0 \\
0&k&0&0 \\
\end{array} \right),
\left(\begin{array}{cccc}
0&0&-\omega&0 \\
0&0&0&0 \\
-\omega&0&0&k \\
0&0&k&0 \\
\end{array} \right),
\left(\begin{array}{cccc}
-2\omega&0&0&k \\
0&0&0&0 \\
0&0&0&0 \\
k&0&0&0 \\
\end{array} \right),
\eeqa
pure-gauge field solutions of the form (\ref{quagetransfgravity})
$e_{\gamma\acute{\gamma}}=
k_{\gamma}\xi_{\acute{\gamma}}+k_{\acute{\gamma}}\xi_{\gamma}$ as one can
see from the  relation
\beqa
H_{\alpha\acute{\alpha}}^{~~~\gamma\acute{\gamma}}(k)
(k_{\gamma}\xi_{\acute{\gamma}}+k_{\acute{\gamma}}\xi_{\gamma})=0.
\eeqa
When $\omega^2 - k^2 = 0$, then the rank of the matrix
$
H_{\alpha\acute{\alpha}\gamma\acute{\gamma}}(k)
$
drops and is equal to $rankH\vert_{\omega^2 - k^2 = 0}  =4$.
This leaves us with $10-4=6$ solutions. These are four
pure-gauge solutions (\ref{quagetransfgravity})
and two additional symmetric solutions representing propagating modes:
the helicity states of the graviton
\beqa
e^{(1)}_{\gamma\acute{\gamma}}=
\left(\begin{array}{cccc}
0&0&0&0 \\
0&1&0&0 \\
0&0&-1&0 \\
0&0&0&0 \\
\end{array} \right),~~
e^{(2)}_{\gamma\acute{\gamma}}=
\left(\begin{array}{cccc}
0&0&0&0 \\
0&0&1&0 \\
0&1&0&0 \\
0&0&0&0 \\
\end{array} \right).
\eeqa
Thus the
general solution of the equation on mass-shell is:
$$
e_{\gamma\acute{\gamma}}=\xi_{\gamma}k_{\acute{\gamma}}+\xi_{\acute{\gamma}}k_{\gamma}
+c_{1}e^{(1)}_{\gamma\acute{\gamma}}+c_{2}e^{(2)}_{\gamma\acute{\gamma}},
$$
where $c_{1},c_{2}$ are arbitrary constants.
We see that the number of propagating modes  is
$$
rankH\vert_{\omega^2 - k^2 \ne 0}-
rankH\vert_{\omega^2 - k^2 = 0}=6-4=2,
$$
as it should be.

\subsection{\it Antisymmetric Tensor Gauge Field }

The antisymmetric second-rank tensor field $B_{\mu\nu}$ is governed by the
equation \cite{Kalb:1974yc,Cremmer:1973mg,Townsend:1976vc,Nambu:1975ba}:
\be\label{antisymmetric}
\partial^{2} B_{\nu\lambda}
-\partial_{\nu} \partial^{\mu}  B_{\mu\lambda} +
\partial_{\lambda} \partial^{\mu}  B_{\mu\nu}=0,
\ee
which is invariant with respect to  the gauge transformations
\beqa
\label{kalbramond}
\delta
B_{\mu\lambda}=\partial_{\mu}\eta_{\lambda} - \partial_{\lambda}\eta_{\mu},
\eeqa
which preserve the symmetry properties of $B_{\mu\nu}$. The corresponding
matrix operator is:
\beqa\label{antisymmetric}
H_{\alpha\acute{\alpha}\gamma\acute{\gamma}}(k)=&-&{1\over 2}
(\eta_{\alpha\gamma}\eta_{\acute{\alpha}\acute{\gamma}}-
\eta_{\alpha\acute{\gamma}}\eta_{\acute{\alpha}  \gamma})
k^{2}-\nn\\
&-&{1\over 2}(\eta_{\alpha\acute{\gamma}}k_{\gamma}k_{\acute{\alpha}}-
\eta_{\acute{\alpha}\acute{\gamma}}k_{\alpha}k_{\gamma}+
\eta_{\acute{\alpha}\gamma}k_{\alpha}k_{\acute{\gamma}}
-\eta_{\alpha\gamma}k_{\acute{\alpha}}k_{\acute{\gamma}})
\eeqa
and is $6 \times 6$ matrix in four-dimensional space-time with the
property $H_{\alpha\acute{\alpha}\gamma\acute{\gamma}}=-
H_{\acute{\alpha}\alpha\gamma\acute{\gamma}} =-
H_{\alpha\acute{\alpha}\acute{\gamma}\gamma}=
H_{\gamma\acute{\gamma}\alpha\acute{\alpha}}$ and is presented in
Appendix B.

If $\omega^2 - k^2 \neq 0$, the rank of the 6-dimensional
matrix
$
H_{\alpha\acute{\alpha}}^{~~~\gamma\acute{\gamma}}(k)
$
is equal to $rankH\vert_{\omega^2 - k^2 \ne 0}=3$
and the number of independent solutions is $6-3=3$.
These three antisymmetric solutions are pure-gauge fields. Indeed,
in  the coordinate system
$k^{\gamma}=(\omega,0,0,k)$
one can find the following three solutions:
\beqa
e_{\gamma\acute{\gamma}}=
\left(\begin{array}{cccc}
0&0&\omega&0 \\
0&0&0&0 \\
-\omega&0&0&k \\
0&0&-k&0 \\
\end{array} \right),
\left(\begin{array}{cccc}
0&\omega&0&0 \\
-\omega&0&0&k \\
0&0&0&0 \\
0&-k&0&0 \\
\end{array} \right),
\left(\begin{array}{cccc}
0&0&0&1 \\
0&0&0&0 \\
0&0&0&0 \\
-1&0&0&0 \\
\end{array} \right),
\eeqa
pure-gauge fields  of the form (\ref{kalbramond})
$e_{\gamma\acute{\gamma}}=
k_{\gamma}\eta_{\acute{\gamma}}- k_{\acute{\gamma}}\eta_{\gamma}$, as one can
see from the  relation
\beqa
H_{\alpha\acute{\alpha}}^{~~~\gamma\acute{\gamma}}(k)
(k_{\gamma}\eta_{\acute{\gamma}}- k_{\acute{\gamma}}\eta_{\gamma})=0.
\eeqa
When $\omega^2 - k^2 = 0$, then the rank of the matrix
$
H_{\alpha\acute{\alpha}\gamma\acute{\gamma}}(k)
$
drops and is equal to $rankH\vert_{\omega^2 - k^2 = 0}  =2$.
This leaves us with $6-2=4$ solutions. These are three
pure-gauge solutions (\ref{kalbramond})
and additional antisymmetric solution representing the propagating mode:
the helicity zero state
\beqa
e^{(A)}_{\gamma\acute{\gamma}}=
\left(\begin{array}{cccc}
0&0&0&0 \\
0&0&1&0 \\
0&-1&0&0 \\
0&0&0&0 \\
\end{array} \right).
\eeqa
Thus on mass-shell the general solution of the equation  is:
$$
e_{\gamma\acute{\gamma}}=k_{\gamma} \eta_{\acute{\gamma}}-
k_{\acute{\gamma}}\eta_{\gamma}
+c_3~ e^{(A)}_{\gamma\acute{\gamma}} ,
$$
where $c_3$ is arbitrary constant.
We see that the number of propagating modes  is
$$
rankH\vert_{\omega^2 - k^2 \ne 0}-
rankH\vert_{\omega^2 - k^2 = 0}=3-2=1.
$$
After this parenthetic discussion we shall
turn to the tensor gauge theory.

\subsection{{\it Non-Abelian Tensor Gauge Field}}

Now we are ready to consider the equation   (\ref{basicequation})
for the tensor gauge field $A_{\mu\lambda}$
with the matrix operator (\ref{quadraticform}), which
in four-dimensional space-time  is a $16 \times 16$ matrix.
In the reference frame, where
$k^{\gamma}=(\omega,0,0,k)$, it has the form presented in the Appendix B.

If $\omega^2 - k^2 \neq 0$, the rank of the 16-dimensional
matrix
$
H_{\alpha\acute{\alpha}}^{~~~\gamma\acute{\gamma}}(k)
$
is equal to $rankH\vert_{\omega^2 - k^2 \ne 0}=9$
and the number of linearly independent solutions is $16-9=7$.
These seven solutions are:
\beqa\label{explicitgaugesolutions}
 e_{\gamma\acute{\gamma}}=
\left(\begin{array}{cccc}
-\omega^{2}&0&0&0 \\
0&0&0&0 \\
0&0&0&0 \\
0&0&0&k^{2} \\
\end{array} \right)
&,&\left(\begin{array}{cccc}
\omega&0&0&0 \\
0&0&0&0 \\
0&0&0&0 \\
k&0&0&0 \\
\end{array} \right),
\left(\begin{array}{cccc}
0&\omega&0&0 \\
0&0&0&0 \\
0&0&0&0 \\
0&k&0&0 \\
\end{array} \right),\left(\begin{array}{cccc}
0&0&\omega&0 \\
0&0&0&0 \\
0&0&0&0 \\
0&0&k&0 \\
\end{array} \right),\nn\\ \left(\begin{array}{cccc}
\omega&0&0&k \\
0&0&0&0 \\
0&0&0&0 \\
0&0&0&0 \\
\end{array} \right)&,&\left(\begin{array}{cccc}
0&0&0&0 \\
\omega&0&0&k \\
0&0&0&0 \\
0&0&0&0 \\
\end{array} \right),\left(\begin{array}{cccc}
0&0&0&0 \\
0&0&0&0 \\
\omega&0&0&k \\
0&0&0&0 \\
\end{array} \right)
\eeqa
pure-gauge tensor potentials of the form (\ref{dualgaugetransfromations})
\be\label{puregaugepotentials}
e_{\gamma\acute{\gamma}}=
k_{\gamma}\xi_{\acute{\gamma}}+k_{\acute{\gamma}}\eta_{\gamma},
\ee
as one can get convinced from the relation
\beqa
H_{\alpha\acute{\alpha}}^{~~~\gamma\acute{\gamma}}(k)
(k_{\gamma}\xi_{\acute{\gamma}}+k_{\acute{\gamma}}\eta_{\gamma})=0,
\eeqa
which follows from the gauge invariance of the action
and can be checked also explicitly.

When $\omega^2 - k^2 = 0$, then the rank of the matrix
$
H_{\alpha\acute{\alpha}\gamma\acute{\gamma}}(k)
$
drops and is equal to $rankH\vert_{\omega^2 - k^2 = 0}  =6$.
This leaves us with $16-6=10$ solutions. These are 7 solutions,
the pure-gauge potentials (\ref{explicitgaugesolutions}), (\ref{puregaugepotentials})
and new three solutions representing propagating modes:
\beqa\label{physicalmodes}
e_{\gamma\acute{\gamma}}^{(1)}=\left(\begin{array}{cccc}
0&0&0&0 \\
0&-1&0&0 \\
0&0&1&0 \\
0&0&0&0 \\
\end{array} \right),~~
e_{\gamma\acute{\gamma}}^{(2)}=
\left(\begin{array}{cccc}
0&0&0&0 \\
0&0&1&0 \\
0&1&0&0 \\
0&0&0&0 \\
\end{array} \right),~~
e_{\gamma\acute{\gamma}}^{A}=
\left(\begin{array}{cccc}
0&0&0&0 \\
0&0&1&0 \\
0&-1&0&0 \\
0&0&0&0 \\
\end{array} \right)
\eeqa
Thus the
general solution of the equation on mass-shell is:
\be
e_{\gamma\acute{\gamma}}=\xi_{\acute{\gamma}}k_{\gamma} +
\eta_{\gamma} k_{\acute{\gamma}}+
c_{1}e^{(1)}_{\gamma\acute{\gamma}}+c_{2}e^{(2)}_{\gamma\acute{\gamma}}
+c_3 e^{(A)}_{\gamma\acute{\gamma}},
\ee
where $c_{1},c_{2}, c_{3}$ are arbitrary constants.
We see that the number of  propagating modes  is three
$$
rankH\vert_{\omega^2 - k^2 \ne 0}-
rankH\vert_{\omega^2 - k^2 = 0}=9-6=3.
$$
These are propagating modes of {\it helicity-two and helicity-zero
$\lambda = \pm 2, 0$ charged gauge bosons}
\cite{Savvidy:2005fi,Savvidy:2005zm,Savvidy:2005ki}.
Indeed, if we make a rotation around the z-axis
\beqa
\Lambda^{\alpha}_{~\beta}=\left(\begin{array}{cccc}
1&0&0&0 \\
0&\cos{\theta}&-\sin{\theta}&0 \\
0&\sin{\theta}&\cos{\theta}&0 \\
0&0&0&1 \\
\end{array} \right), \nn
\eeqa
we shall get
\beqa
e^{(1)'}=\Lambda e^{(1)}\Lambda^{T}=\left(\begin{array}{cccc}
0&0&0&0 \\
0&-\cos{2\theta}&-\sin{2\theta}&0 \\
0&-\sin{2\theta}&\cos{2\theta}&0 \\
0&0&0&0 \\
\end{array} \right),  ~~~
e^{(2)'}=\Lambda e^{(2)}\Lambda^{T}=\left(\begin{array}{cccc}
0&0&0&0 \\
0&-\sin{2\theta}&\cos{2\theta}&0 \\
0&\cos{2\theta}&\sin{2\theta}&0 \\
0&0&0&0 \\
\end{array} \right). \nn
\eeqa
Therefore the first two solutions describe helicity $\lambda=\pm2$ states.
On the other hand, the third, antisymmetric solution remains invariant
under a Lorentz transformation, therefore it describes helicity-zero state.

This result can also be derived from the consideration of the equations of motion
for the symmetric and antisymmetric parts of the tensor gauge
field $A^{a}_{\mu\lambda}$, as it was done in
\cite{Savvidy:2005fi,Savvidy:2005zm,Savvidy:2005ki}.
Indeed,  one can observe that for
the symmetric part of the tensor gauge fields $A^{a}_{\nu\lambda}$
the equation reduces to the previously studied free equation in gravity
(\ref{einstein}), which describes the propagation of
massless tensor gauge bosons with two physical polarizations: the
$\lambda= \pm 2$ helicity states. For the antisymmetric part of the
tensor gauge fields $A^{a}_{\nu\lambda}$
the equation reduces to the equation
(\ref{kalbramond}) which describes the propagation of
massless boson  with one physical polarization: the
$\lambda= 0$ helicity state.

\section{\it Interaction of Currents }

The interaction amplitude between two tensor currents caused by the exchange
of these tensor gauge bosons can be found from (\ref{freeequations1}),
(\ref{mainequation}) and (\ref{tensorfieldequation}) and has the following
form \cite{Savvidy:2005ki}
\be
J^{'}_{\mu\lambda} ~\Delta^{\mu\lambda\nu\rho}~J_{\nu\rho}
\ee
where the propagator $\Delta^{ab}_{\mu\lambda\nu\rho}$ is
\be\label{interamplitude}
\Delta^{ab}_{\mu\lambda\nu\rho}= \delta^{ab}{\eta_{\mu\nu} \eta_{\lambda\rho} -
{1\over 2} \eta_{\mu\lambda} \eta_{\nu\rho}
\over \omega^2 - k^2},
\ee
therefore
\be
J^{'}_{\mu\lambda} ~\Delta^{\mu\lambda\nu\rho}~J_{\nu\rho} =
J^{'}_{\mu\lambda} ~{1\over \omega^2 - k^2  } ~J^{\mu\lambda}-
{1\over 2} J^{'~\mu}_{\mu} ~ {1\over \omega^2 - k^2  } ~J^{~\lambda}_{\lambda}.
\ee
We shall evaluate the first term in the interaction amplitude. This gives
\beqa
  J^{'}_{\mu\lambda} ~{1\over \omega^2 - k^2  }~J^{\mu\lambda}= {1\over \omega^2 - k^2  }
  \{J^{'}_{00 }J_{ 00} -J^{'}_{01 }J_{01 }-J^{'}_{02 }J_{02 }-J^{'}_{03 }J_{03 }
  -J^{'}_{10 }J_{10 }-J^{'}_{20 }J_{20 }-J^{'}_{30 }J_{30 }\nn\\
  +J^{'}_{11 }J_{11 }+J^{'}_{22 }J_{22 }+J^{'}_{33 }J_{33 }
  +J^{'}_{12 }J_{12 }+J^{'}_{21 }J_{21 }+J^{'}_{13 }J_{13 }
  +J^{'}_{31 }J_{31 }+J^{'}_{23 }J_{23 }+J^{'}_{32 }J_{ 32}\}. \nn
\eeqa
Taking $k^{\mu} =(\omega,0,0,k)$ and using the conservation of the
current (\ref{divergentfirst}) expressed in the momentum space
\beqa
k^{\mu}J_{\mu\lambda}= 0, ~~~~~~\omega J_{0\lambda}= - k~ J_{3\lambda}\nn
\eeqa
we shall get
\beqa
{1\over \omega^2 - k^2  }
  [ (1-{\omega^2 \over k^2})J^{'}_{00 }J_{ 00}
    -(1-{\omega^2 \over k^2})J^{'}_{01 }J_{01 }-
    (1-{\omega^2 \over k^2})J^{'}_{02 }J_{02 }-
    (1-{\omega^2 \over k^2})J^{'}_{03 }J_{03 }\nn\\
  -J^{'}_{10 }J_{10 }-J^{'}_{20 }J_{20 }
  +J^{'}_{11 }J_{11 }+J^{'}_{22 }J_{22 }
  +J^{'}_{12 }J_{12 }+J^{'}_{21 }J_{21 }+J^{'}_{13 }J_{13 }
  +J^{'}_{23 }J_{23 }] .\nn
\eeqa
Now using the second conservation law (\ref{divergentsecond}) in momentum space
\beqa
k^{\lambda}J_{\mu\lambda}= 0, ~~~~~~\omega J_{\mu0}= - k~ J_{\mu3}\nn
\eeqa
we arrive at
\beqa
 {1\over \omega^2 - k^2  }
  [ (1-{\omega^2 \over k^2})J^{'}_{00 }J_{ 00}
    -(1-{\omega^2 \over k^2})J^{'}_{01 }J_{01 }-
    (1-{\omega^2 \over k^2})J^{'}_{02 }J_{02 }-\nn\\
    -(1-{\omega^2 \over k^2})J^{'}_{10 }J_{10 }-
     (1-{\omega^2 \over k^2})J^{'}_{20 }J_{20 }
     - (1-{\omega^2 \over k^2})({\omega^2 \over k^2}) J^{'}_{00 }J_{00 }] +\nn\\
  +{1\over \omega^2 - k^2  }[J^{'}_{11 }J_{11 }+J^{'}_{22 }J_{22 }
  +J^{'}_{12 }J_{12 }+J^{'}_{21 }J_{21 }]\nn
\eeqa
and after simple algebra at
\beqa
  -{1\over k^2  }
  [(1-{\omega^2 \over k^2})J^{'}_{00 }J_{ 00} - J^{'}_{01 }J_{01 }-
     J^{'}_{02 }J_{02 }
     - J^{'}_{10 }J_{10 }-
     J^{'}_{20 }J_{20 }]+\nn\\
  +{1\over \omega^2 - k^2  }[J^{'}_{11 }J_{11 }+J^{'}_{22 }J_{22 }
  +J^{'}_{12 }J_{12 }+J^{'}_{21 }J_{21 }].\nn
\eeqa
Evaluating the second term in the interaction amplitude (\ref{interamplitude})
in the same manner as above, we shall finally get for the total amplitude:
\beqa
 -{1\over k^2  }
  [(1-{\omega^2 \over k^2})J^{'}_{00 }J_{ 00} - J^{'}_{01 }J_{01 }-
     J^{'}_{02 }J_{02 }
     - J^{'}_{10 }J_{10 }-
     J^{'}_{20 }J_{20 }]+\nn\\
  +{1\over \omega^2 - k^2  }[{1\over 2}(J^{'}_{11 }- J^{'}_{22 })(J_{11 }- J_{22 })
  +J^{'}_{12 }J_{12 }+J^{'}_{21 }J_{21 }].
\eeqa
For the instantaneous term we get
\be
-{1\over k^2  }
  [(1-{\omega^2 \over k^2})J^{'}_{00 }J_{ 00} - J^{'}_{01 }J_{01 }-
     J^{'}_{02 }J_{02 }
     - J^{'}_{10 }J_{10 }-
     J^{'}_{20 }J_{20 }]
\ee
and for the retarded term ($J_{12 } \neq J_{21 }$)
\be
{1\over \omega^2 - k^2  }[{1\over 2}(J^{'}_{11 }- J^{'}_{22 })(J_{11 }- J_{22 })
  +J^{'}_{12 }J_{12 }+J^{'}_{21 }J_{21 }].
\ee
The retarded term represents a sum of three independent products
\beqa
&+& {1\over 4}[J^{'}_{11 }- J^{'}_{22 } +i(J^{'}_{12 } +J^{'}_{21 } )]
[J_{11 }- J_{22 } -i(J_{12 } +J_{21 } )]+\nn\\
&+& {1\over 4}[J^{'}_{11 }- J^{'}_{22 } -i(J^{'}_{12 } +J^{'}_{21 } )]
[J_{11 }- J_{22 } + i(J_{12 } +J_{21 } )]+\nn\\
&+& {1\over 2} (J^{'}_{12 } -J^{'}_{21 } )
(J^{'}_{12 } -J^{'}_{21 } )
\eeqa
or polarizations corresponding to the helicities $\lambda = \pm 2 ,0$.
Thus all negative-norm states are excluded from the spectrum of the
second-rank  tensor gauge field $A_{\mu\lambda}$, due
to the gauge invariance of the theory and
we come to the conclusion that the theory
does indeed respect unitarity at the free level.

\section{{\it Appendix A.  Bianchi identity}}

The non-Abelian tensor fields $A^{a}_{\mu\lambda_{1}...
\lambda_{s}}$ can be seen as appearing in the expansion of the
extended gauge field ${\cal A}_{\mu}(x,e)$ over the unit tangent
vector $e_{\lambda}$ \cite{Savvidy:2005fi,Savvidy:2005zm,Savvidy:2005ki}:
\beqa
{\cal A}_{\mu}(x,e)=\sum_{s=0}^{\infty} {1\over s!} ~A^{a}_{\mu\lambda_{1}...\nn
\lambda_{s}}(x)~L^{a}e_{\lambda_{1}}...e_{\lambda_{s}}.
\eeqa
and the extended field strength tensor can be defined in terms of the extended gauge
field ${\cal A}_{\mu}(x,e)$ as follows:
\beqa
{\cal G}_{\mu\nu}(x,e)=\partial_{\mu}{\cal A}_{\nu}(x,e)-\partial_{\nu}{\cal A}_{\mu}(x,e)-
ig [{\cal A}_{\mu}(x,e),{\cal A}_{\nu}(x,e)].\nn
\eeqa
Defining the extended covariant derivative:
${\cal D}_{\mu}=\partial_{\mu}-ig{\cal A}_{\mu},$
one can get \cite{Savvidy:2005ki}:
\beqa
\label{covariantcommutator}
[{\cal D}_{\mu},{\cal D}_{\nu}]=[\partial_{\mu}-ig{\cal A}_{\mu},
\partial_{\nu}-ig{\cal A}_{\nu}]=
-ig {\cal G}_{\mu\nu}.
\eeqa
The operators ${\cal D}_{\mu},{\cal D}_{\nu},{\cal D}_{\lambda}$ obey Jacobi identity:
$$
[{\cal D}_{\mu},[{\cal D}_{\nu},{\cal D}_{\lambda}]]+
[{\cal D}_{\nu},[{\cal D}_{\lambda},{\cal D}_{\mu}]]+
[{\cal D}_{\lambda},[{\cal D}_{\mu},{\cal D}_{\nu}]]  =0,
$$
which with the aid of (\ref{covariantcommutator}) is transformed into the
generalized Bianchi identity
\beqa
\label{Bianchiextended}
[{\cal D}_{\mu},{\cal G}_{\nu\lambda}]+[{\cal D}_{\nu},{\cal G}_{\lambda\mu}]+
[{\cal D}_{\lambda},{\cal G}_{\mu\nu}]  =0.
\eeqa
Let us now expand equation (\ref{Bianchiextended}) over $e_{\rho}$
up to linear terms. We have,
$$
[\partial_{\mu}-igA_{\mu}-igA_{\mu\rho}e^{\rho},~G_{\nu\lambda}+G_{\nu\lambda,\rho}e^{\rho}]
+cyc.perm. + O(e^{2})=0
$$
In zero order the above equation gives the standard Bianchi identity in YM theory:
$$
[D_{\mu},G_{\nu\lambda}]+[D_{\nu},G_{\lambda\mu}]+
[D_{\lambda},G_{\mu\nu}]=0,
$$
where $D_{\mu}=\partial_{\mu}-igA_{\mu}$.
The linear term in $e_{\rho}$ gives:
\be\label{newbianchi}
[D_{\mu},G_{\nu\lambda,\rho}]-ig[A_{\mu\rho},G_{\nu\lambda}]+
[D_{\nu},G_{\lambda\mu,\rho}]-ig[A_{\nu\rho},G_{\lambda\mu}]+
[D_{\lambda},G_{\mu\nu,\rho}]-ig[A_{\lambda\rho},G_{\mu\nu}]=0
\ee
Using explicit form of the operators $D_{\mu}$, $G_{\mu\nu}$ and $G_{\mu\nu,\lambda}$
one can independently check the last identity and get convinced that it holds.
Now, if we expand the above equation over g, the zeroth order gives the Bianchi identity
for the field strength tensor $F_{\nu\lambda,\rho}$:
\be
\label{Bianchiidentity1}
\partial_{\mu}F_{\nu\lambda,\rho}+\partial_{\nu}F_{\lambda\mu,\rho}
+\partial_{\lambda}F_{\mu\nu,\rho}=0
\ee
These equations impose tight restrictions on the source
currents and hence on the nature of interactions.

\section{\it Appendix B}

The matrix operator in gravity (\ref{eqsmatrixgravity}) is of dimension
$10 \times 10$ and has the following form
\be
\begin{array}{cccccccccc}
    0&0&0&0&-{k^2}&0&0&-{k^2}&0&0 \\
    0&-\frac{{k^2}}{2}&0&0&0&0&-\frac{k  w}{2}&0&0&0 \\
    0&0&-\frac{{k^2}}{2}&0&0&0&0&0&-\frac{k  w}{2}&0 \\
    0&0&0&0&k  w&0&0&k  w&0&0 \\
    -{k^2}&0&0&-k  w&0&0&0&{k^2}-{w^2}&0&-{w^2} \\
    0&0&0&0&0&\frac{1}{2}  (-{k^2}+{w^2})&0&0&0&0 \\
    0&\frac{k  w}{2}&0&0&0&0&\frac{{w^2}}{2}&0&0&0 \\
    -{k^2}&0&0&-k  w&{k^2}-{w^2}&0&0&0&0&-{w^2} \\
    0&0&\frac{k  w}{2}&0&0&0&0&0&\frac{{w^2}}{2}&0 \\
    0&0&0&0&-{w^2}&0&0&-{w^2}&0&0
  \end{array}
\ee
and for antisymmetric tensor field (\ref{antisymmetric}) it is of dimension
$6 \times 6$
\be
\begin{array}{cccccc}
    -\frac{{k^2}}{2}&0&0&0&\frac{k  w}{2}&0 \\
    0&-\frac{{k^2}}{2}&0&0&0&\frac{k  w}{2} \\
    0&0&0&0&0&0 \\
    0&0&0&\frac{1}{2}  (-{k^2}+{w^2})&0&0 \\
    -\frac{k  w}{2}&0&0&0&\frac{{w^2}}{2}&0 \\
    0&-\frac{k  w}{2}&0&0&0&\frac{{w^2}}{2} .
\end{array}
\ee
The matrix operator for non-Abelian tensor gauge theory (\ref{quadraticform})
is of dimension
$16 \times 16$ and has the following explicit form:
{\fontsize{2}{2.2}
\be
\begin{array}{cccccccccccccccc}
    0&0&0&0&0&-\frac{{k^2}}{2}&0&0&0&0&-\frac{{k^2}}{2}&0&0&0&0&0 \\
    0&-{k^2}&0&0&\frac{{k^2}}{2}&0&0&\frac{k \omega}{2}&0&0&0&0&0&-k \omega&0&0 \\
    0&0&-{k^2}&0&0&0&0&0&\frac{{k^2}}{2}&0&0&\frac{k \omega}{2}&0&0&-k \omega&0 \\
    0&0&0&0&0&\frac{k \omega}{2}&0&0&0&0&\frac{k \omega}{2}&0&0&0&0&0 \\
    0&\frac{{k^2}}{2}&0&0&-{k^2}&0&0&-k \omega&0&0&0&0&0&\frac{k \omega}{2}&0&0 \\
    -\frac{{k^2}}{2}&0&0&-\frac{k \omega}{2}&0&0&0&0&0&0&\frac{1}{2} ({k^2}-{\omega^2})&0&-\frac{k \omega}{2}&0&0&-\frac{{\omega^2}}{2} \\
    0&0&0&0&0&0&-{k^2}+{\omega^2}&0&0&\frac{1}{2} ({k^2}-{\omega^2})&0&0&0&0&0&0 \\
    0&-\frac{k \omega}{2}&0&0&k \omega&0&0&{\omega^2}&0&0&0&0&0&-\frac{{\omega^2}}{2}&0&0 \\
    0&0&\frac{{k^2}}{2}&0&0&0&0&0&-{k^2}&0&0&-k \omega&0&0&\frac{k \omega}{2}&0 \\
    0&0&0&0&0&0&\frac{1}{2} ({k^2}-{\omega^2})&0&0&-{k^2}+{\omega^2}&0&0&0&0&0&0 \\
    -\frac{{k^2}}{2}&0&0&-\frac{k \omega}{2}&0&\frac{1}{2} ({k^2}-{\omega^2})&0&0&0&0&0&0&-\frac{k \omega}{2}&0&0&-\frac{{\omega^2}}{2} \\
    0&0&-\frac{k \omega}{2}&0&0&0&0&0&k \omega&0&0&{\omega^2}&0&0&-\frac{{\omega^2}}{2}&0 \\
    0&0&0&0&0&\frac{k \omega}{2}&0&0&0&0&\frac{k \omega}{2}&0&0&0&0&0 \\
    0&k \omega&0&0&-\frac{k \omega}{2}&0&0&-\frac{{\omega^2}}{2}&0&0&0&0&0&{\omega^2}&0&0 \\
    0&0&k \omega&0&0&0&0&0&-\frac{k \omega}{2}&0&0&-\frac{{\omega^2}}{2}&0&0&{\omega^2}&0 \\
    0&0&0&0&0&-\frac{{\omega^2}}{2}&0&0&0&0&-\frac{{\omega^2}}{2}&0&0&0&0&0
  \end{array}
\ee
}
and allows to calculate its rank as a function of momenta.

\vfill
\end{document}